\documentclass[pdftex,floatfix,pre,aps,amsfonts,twocolumn,cite]{revtex4}

\usepackage{amsmath,graphicx,xcolor,bbm,soul}

\usepackage{verbatim}

\def\ta{\tau_{\rm e}}
\def\P{\mathbbm P}
\def\E{\mathbbm E}
\def\V{\mathbbm V}
\def\1{\mathbbm{1}}
\def\mD{\mathfrak D}

\begin{document}

\date{\today}

\title{Emergence of step flow from atomistic scheme of epitaxial growth in 1+1 dimensions}

\author{Jianfeng Lu$^{1,2}$, Jian-Guo Liu$^1$, and Dionisios Margetis$^3$}

\affiliation{${}^1$Department of Mathematics and Department of Physics, Duke University,
Durham, North Carolina 27708, USA\\
${}^2$Department of Chemistry, Duke University, Durham, North Carolina 27708, USA\\
${}^3$Department of Mathematics, and Institute for Physical Science and Technology, and Center for Scientific
Computation and Mathematical Modeling, University of Maryland, College Park,
Maryland 20742, USA}

\begin{abstract}
The Burton-Cabrera-Frank (BCF) model for the flow of line defects (steps) on crystal surfaces has offered useful insights into nanostructure evolution. This model has rested
on phenomenological grounds. Our goal is to show via scaling arguments the {\em emergence} of
the BCF theory for non-interacting steps {\em from a stochastic atomistic scheme} of a kinetic restricted  solid-on-solid model in {\em one spatial dimension}. Our main assumptions are: adsorbed atoms (adatoms) form a dilute system, and elastic effects of the crystal lattice are absent.
The step edge is treated as a front that propagates via probabilistic rules for atom attachment and detachment at the step.
We formally derive a quasistatic step flow description by averaging out the stochastic scheme when terrace diffusion, adatom desorption and
deposition from above are present.

\vskip5pt

\noindent PACS numbers: 81.15.Aa, 68.43.Jk, 47.11.St
\end{abstract}

\maketitle

\section{Introduction}
\label{sec:intro}

The design and fabrication of optoelectronic devices rely on understanding how crystalline features evolve across several length scales, from a few nanometers to hundreds of microns.
At low enough temperatures, below the roughening transition, crystal surface structures evolve through the collective motion of line defects, steps~\cite{PimpinelliVillain99,JW99,Misbah10}. The motion of individual steps is a mesoscale phenomenon: On the one hand, it manifests defects of {\em atomic} size; on the other hand, steps appear to move in a {\em continuum} fashion by exchanging mass with nanoscale regions, terraces.
For the description of crystal surface dynamics in a wide range of length and time scales, it is thus useful to explore the validity and applicability of mesoscale models for step flow.
These models capture atomistic features in the direction vertical to the high-symmetry plane of the crystal, while retaining the advantages of continuum theories in
the lateral directions.

Such a hybrid approach is the Burton-Cabrera-Frank (BCF) model~\cite{BCF51}; for reviews, see, e.g.,~\cite{PimpinelliVillain99,JW99}.
In this model, step edges are represented by moving smooth curves, which are boundaries of terraces. The step motion is mediated by the continuous diffusion of
adsorbed atoms (adatoms). A typical BCF-type description consists of the following~\cite{PimpinelliVillain99,JW99}: (i) a step velocity law; (ii) the diffusion
equation for the adatom density on each terrace; and (iii) a near-equilibrium, linear kinetic relation that involves the adatom flux normal to the step edge and forms an extension of the boundary condition for the adatom density in~\cite{BCF51}.
The motion laws for steps have been conceived phenomenologically by the principles of mass conservation and local thermodynamic equilibrium.
The connection of this mesoscale picture to fundamental atomistic processes is not adequately understood.

In this paper, we develop a stochastic scheme adopted from a kinetic restricted solid-on-solid (SOS) model~\cite{Weeks,PatroneEM14} for the hopping of atoms on a stepped surface in 1+1 dimensions (one spatial dimension, 1D, plus time)  in the absence of elastic effects.
We derive the BCF description for the flow of steps as a {\em scaling limit} of averaged equations of the atomistic model.
First, we analyze an epitaxial system with a single step in the presence of external material deposition and desorption;
and then extend our analysis to many steps.
We assume that the adatoms are non-interacting and form
a {\em dilute} system; thus, on average, only a small number of adatoms occupy each lattice site at any given time. This diluteness has been observed experimentally~\cite{Zhu98},
and simplifies the atomistic laws.

Our present approach is inspired by recent efforts to shed light on the nature of the BCF theory~\cite{PatroneM14,PatroneEM14}; see also~\cite{Zhao05,Ackerman11,Saum09}. It is tempting to explore whether the BCF model can be interpreted as the universal, in some appropriate sense, limit
of atomistic processes. Adopting a line of investigation that favors this view, we invoke
basic mechanisms of atomistic motion in the presence of steps; these include generic local rules for the atom attachment/detachment at a step edge.
Our hypotheses lead to a linear kinetic relation between the mesoscale adatom flux and the adatom density  in the presence of a step-edge energy barrier on both sides of the step~\cite{JW99,ES}. We also discuss the case when such a barrier is absent.

This study is motivated  by the broader question how to develop mesoscale models for crystal defects. 
A long-term objective is to construct by purely atomistic principles
mesoscale theories for kinetic regimes far from thermodynamic equilibrium (for
related models, see~\cite{Voronkov,Caflischetal99,Filimonov04,Balykov05}).

Several past works~\cite{PatroneM14,PatroneEM14,Zhao05,Ackerman11,Saum09,Zangwill92} with a similar perspective should be mentioned.
In~\cite{PatroneM14,PatroneEM14}, the starting point is a master equation for the probabilities of finding a one-step system in atomistic configurations
characterized by the total number of adatoms
and their positions on a one-dimensional lattice.
The mesoscale motion of the step comes from the ensemble average of its microscale position. In this setting, the entire BCF-type description
emerges as the low-density limit of the adatom system~\cite{PatroneM14}. This formalism is not directly extensible to two spatial dimensions (2D). In~\cite{Zhao05}, the authors connect atomistic rates to BCF-type parameters via balancing out discrete and continuum fluxes at the step edge in 1D, without invoking a stochastic scheme or describing the effect of noise; their results are compatible with ours. On the other hand, the studies in~\cite{Ackerman11,Saum09} concern geometries in 2D with focus
on more particular aspects of step flow. For example, in~\cite{Ackerman11} the step position is held fixed;
and in~\cite{Saum09} only numerical comparisons of kinetic Monte Carlo (KMC) simulations to the BCF model are pursued. Notably, in~\cite{Zangwill92} the authors start with a 2D master equation and reduce it to a Langevin-type description
for continuous-in-time height columns by retaining discreteness in the lateral directions. We believe that a direct comparison of this last approach to the BCF theory is not compelling. Other, related yet different in perspective, works focus on characterizing near-equilibrium growth conditions on semiconductor surfaces~\cite{Johnson96,Tersoff97}.

Our derivation of the BCF limit in 1D in this paper differs from the analysis of~\cite{PatroneM14,PatroneEM14} in several interrelated aspects. First, here we apply the hypothesis of a dilute adatom system, whereas in~\cite{PatroneM14} the diluteness results as a special case. Second,
we invoke a stochastic scheme, in contrast to the master-equation approach of~\cite{PatroneM14}. This, along with the diluteness hypothesis, enables us to include richer kinetic effects, namely,
desorption and material deposition from above, and many steps with relative algebraic ease. Third, we introduce the step front position as a stochastic variable whose motion is {\em coupled} with the random number of adatoms per lattice site.

Nucleation is not included in our atomistic model. At low enough temperatures this effect can cause a decrease of the step velocity, enabling deviations from the linear kinetic law of the BCF model~\cite{Shitara93}. Step permeability,
which is usually introduced phenomenologically at the BCF level~\cite{OzdemirZangwill92}, does not directly ensue from our model; an additional atomistic process may be needed to capture this effect (see our remarks in Sec.~\ref{subsec:random}).  

An important aspect of our analysis is the systematic averaging of a stochastic scheme that allows the derivation of BCF-type laws as scaling limits when the lattice spacing approaches zero. We believe that our methodology has not previously been applied in epitaxial growth.

Our analysis reveals how the stochastic noise affects step motion for small  lattice spacing. In fact, we show that, under appropriate scalings of the kinetic rates,  this noise tends to vanish. Hence, the mesoscale step position and adatom density approach their expectation values. In the language of probability theory, the BCF model  with a linear kinetic relation for the mass flux  emerges in a regime where the ``law of large numbers'' applies.

Our starting scheme invokes ideas of a random choice method (``Glimm scheme'') invented for solving certain systems of conservation laws such as those arising in gas
dynamics~\cite{Glimm65,Chorin76}. The main idea is to construct the appropriate solution (say, a shock wave) through
a sequence of operations; these include a sampling scheme by use of a random variable that is uniformly distributed over a fixed interval.
Our approach has a similar flavor but bears particularities tailored to the
physics of epitaxial growth: The time-dependent random variable, $\xi(t)$, that we employ takes discrete values corresponding to the events of advancement, retreat or immobility of the step edge as adatoms attach to the step, detach from it or move otherwise, respectively. These events have prescribed probabilities involving known atomistic rates subject to the principle of detailed balance in the sense of~\cite{Caflischetal99} (see Sec.~\ref{sec:model}).

Our work has limitations. These are mainly due to restricting our
attention to: dilute systems, non-interacting steps, and 1D. In
particular, the possible emergence of force-dipole step-step
interaction~\cite{Marchenko} may require the alteration of the
stochastic scheme to take into account an elastic lattice with
spontaneous stress~\cite{Saito}. We expect that the extension of our
formalism to 2D would have to possibly involve a space-time stochastic noise
driving step fluctuations on the lattice.  In the 2D case, a challenge is that step meandering leads to an effective source of free adatoms on terraces that implies a modification of the concentration entering the BCF model~\cite{Shitara93}. 

The remainder of the paper is organized as follows. In Sec.~\ref{sec:model}, we formulate the discrete stochastic scheme for a single step. In Sec.~\ref{sec:limit},
we formally derive the scaling limit of this scheme.  In Sec.~\ref{sec:discussion}, we discuss implications and
extensions of our analysis, particularly the presence of more than one steps. Section~\ref{sec:conclusion} concludes our work with a summary of our results and an outline of
open problems. Throughout the paper, the expression $Q=\mathcal O(h)$ means that the quantity $Q/h$ is bounded by a constant as a parameter approaches a limit. The bar on top of a symbol for a stochastic variable implies the mean value (expectation) of that variable.

\section{Atomistic scheme with one step}
\label{sec:model}

\begin{figure}
\includegraphics*[width=3.6in,height=0.77in,trim=0.7in 2.6in 0 3.0in]{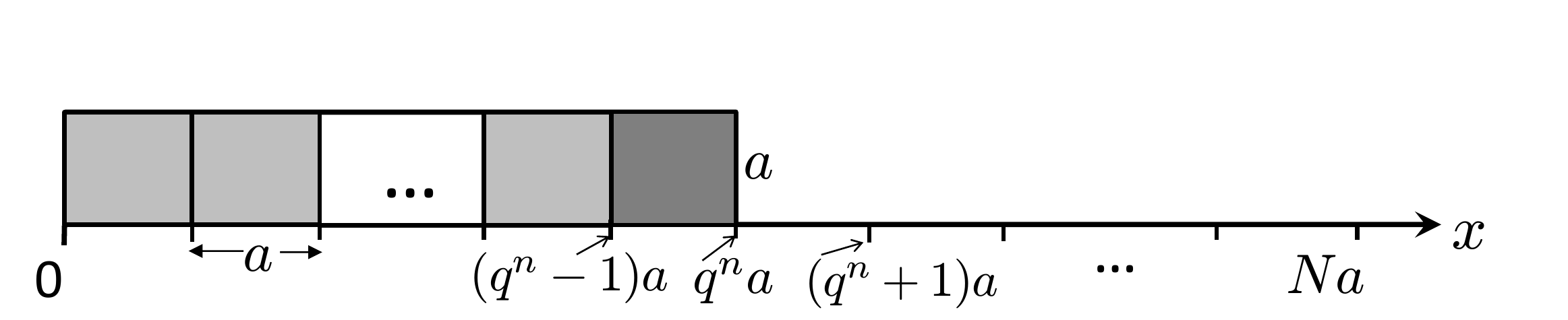}
    \caption{Microscale view of a step edge at time $t=t_n$. The step has height $a$, an atomic length, and lies on a 1D lattice of spacing $a$ and total length $L=Na$,
where $N$ is the total number of lattice sites ($N\gg 1$). The step position is determined by the lattice site $q^n=q(t_n)$ directly to the right of the edge ($q^n=0, 1, \ldots N-1$). The step atoms (grey) of the upper terrace and the edge atom (dark grey) are represented by squares; the atom position is indexed by the left side of each square, as indicated by arrows in the vicinity of the step. The Eulerian coordinate is $x=ja$ ($j=0,\,\ldots N-1$).}\label{fig:geometry}
\end{figure}

The single-step geometry in 1D is shown in Fig.~\ref{fig:geometry}. The step lies on a  lattice of uniform spacing $a$ and length $L=Na$ where $N\gg 1$.
Since $L$ constitutes a natural length of the BCF setting, we set $L=1$; thus, $a=1/N\ll 1$. The step position at time $t$ can be tracked by $q(t)$, an integer-valued Lagrangian coordinate expressing the number of the lattice site located immediately to the right of the step edge [$q(t)=0, 1,\ldots, N-1$]; this $q$ is distinct from $j$, the Eulerian coordinate for the lattice site. Hence, the step edge position is determined through the discrete stochastic variable $X(t)=q(t) a$.

We distinguish the edge atom, which has only one in-plane nearest neighbor, from the step atom, which has two in-plane nearest
neighbors, as shown in Fig.~\ref{fig:geometry}. An adatom is a movable particle that is neither an edge atom nor a step atom.

\subsection{Kinetic frame: Assumptions}
\label{subsec:kins}

To prescribe the adatom kinetics relative to the step edge, we apply the following main assumptions~\cite{PatroneEM14}.

\begin{figure}
\includegraphics*[width=4.2in, trim=1.5in 1.9in 0.2in 1.3in]{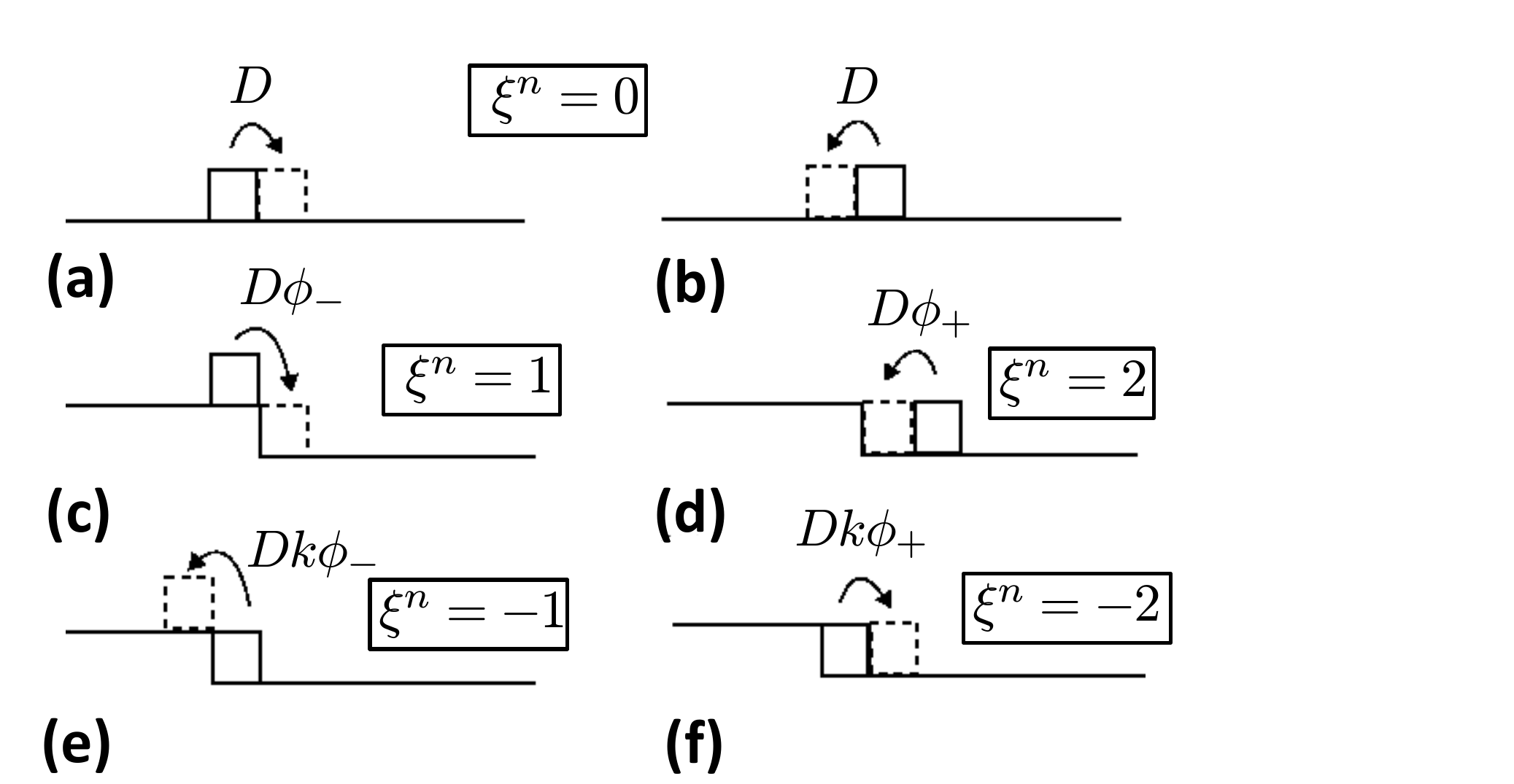}
\caption{Schematic of basic atomistic processes of our model at $t=t_n$ and corresponding values of random variable $\xi(t_n)=\xi^n$ (Sec.~\ref{subsec:random}). {\em Upper panel} [(a), (b)]: Unbiased hopping of adatom with rate $D$ from a lattice site of a terrace to an adjacent site of the same terrace
to the right [(a)] or left [(b)]; the step does not move. 
{\em Middle panel} [(c), (d)]: Attachment of an adatom to the step edge from the upper terrace with rate $D\phi_-$ [(c)], or the lower terrace with rate $D\phi_+$ [(d)]; the step moves to the right by one lattice spacing, $a$.
{\em Lower panel} [(e), (f)]: Detachment of an edge atom from the step to the upper terrace with rate $Dk\phi_-$ [(e)], or the lower terrace with rate $Dk\phi_+$ [(f)]; the step moves to the left by distance $a$.}\label{fig:kinetics}
\end{figure}

\begin{itemize}

\item[(i)] An atom is only allowed to move horizontally, left or right, by one lattice site at any given time.

 \item[(ii)] The adatoms are non-interacting and have low density, i.e., only a small average number of adatoms occupy any lattice site at any given time. Hence, it is unlikely that islands form [see also (vi) below].

\item[(iii)] An adatom can hop from a lattice site to an adjacent site of the same terrace with a probability proportional to the constant rate $D$. This rule gives rise
to the usual, unbiased diffusion process as the result of a random walk [see Figs.~\ref{fig:kinetics}(a), (b)].

\item[(iv)] An adatom from the upper ($-$) or lower ($+$) terrace attaches to the step edge and becomes an edge atom with rate $D\phi_{\mp}$, where the nondimensional $\phi_{\pm}$ ($\phi_{\pm}\le 1$)
accounts for the Ehrlich-Schwoebel barrier~\cite{ES,PimpinelliVillain99,PatroneEM14}; $\phi_{\pm}=e^{-E_{\pm}/T}$, $E_{\pm}>0$, and $T$ is the Boltzmann energy (absolute temperature). As a result, the step edge moves {\em forward} (to the right)
by a distance equal to $a$ [see Figs.~\ref{fig:kinetics}(c), (d)].

\item[(v)] An edge atom can detach from a step, breaking a bond, become an adatom and hop to the upper ($-$) or lower ($+$) terrace with rate $Dk\phi_{\mp}$, where $k=e^{-E_{\rm b}/T}$ and $E_{\rm b}$ is the edge-atom bond energy barrier, $E_{\rm b}>0$. Thus, the step retreats (to the left)  by distance $a$ [Figs.~\ref{fig:kinetics}(e), (f)].

\item[(vi)] A step atom cannot become an adatom, or vice versa.

\item[(vii)] Only adatoms can evaporate from the surface.

\item[(viii)] Atoms deposited on the terrace from above instantly become adatoms.

\end{itemize}

In our atomistic model, steps move only via rules (iv) and (v).
By our choice of kinetic rates at the step edge, detailed
balance is satisfied in the sense of~\cite{Caflischetal99,Otto,Katsoulakis}. This principle implies that at equilibrium the microscale adatom fluxes toward the step edge vanish~\cite{Caflischetal99}. In
particular, by setting $D_{\rm TE}^-=D\phi_-$, $D_{\rm TE}^+=D\phi_+$,
$D_{\rm ET}^-=Dk\phi_-$ and $D_{\rm ET}^+=Dk\phi_+$, we note the relation
$D_{\rm TE}^-D_{\rm ET}^+=D_{\rm TE}^+D_{\rm ET}$.
In the special case of a simple cubic SOS model~\cite{ClarkeVvedensky,Caflischetal99}, it is expected that $D_{\rm TE}^{\pm}=D$ and $D_{\rm ET}^+=D_{\rm ET}^-$; thus, $\phi_+=\phi_-=1$. This plausibly leads to the Dirichlet boundary condition that the adatom density equals an equilibrium density at the step~\cite{PimpinelliVillain99},  which we discuss as a special case in Sec.~\ref{subsec:rates-sc}.

Experimental estimates of $E_{\pm}$, and thus of $\phi_{\pm}$, are outlined in~\cite{PatroneEM14}. For a detailed list of associated values, see Table 6 in~\cite{JW99}. In particular, for Ni(110), one finds $E_- = 0.9$ eV and $E_+ \approx 0$ eV; hence, $\phi_-\ll 1$ and $\phi_+\approx 1$ at 500 K.  
Thus, in a BCF-type description for this system, our analysis predicts a distinct type of boundary condition for each side of the step edge (see Sec.~\ref{subsec:rates-sc}), as expected from past works based on other approaches~\cite{PimpinelliVillain99,PatroneEM14,PatroneM14}.

Atoms are assumed to be deposited on the surface from above with constant flux $f$, which expresses number of atoms per unit time per lattice site,  and can be evaporated with constant rate $\ta^{-1}$ where $\ta$ is a typical evaporation or desorption time.
In addition, we introduce boundary conditions at the fixed points $x=0$ and $x=1$ for definiteness. We consider a steady incoming flux, $f_{\rm in}$, of adatoms from the left boundary, $x=0$. Some of the incoming adatoms attach to the step so that the step moves forward; while some other adatoms
leave the system from the right boundary, $x=1$. Adatoms are not allowed to enter the prescribed spatial domain, $0< x< 1$, from the right boundary or leave it from the left boundary. Other choices of boundary conditions are possible without distorting the step motion laws.  For example, one can alternatively impose screw-periodic boundary conditions in the atomistic description.

\subsection{Stochastic scheme}
\label{subsec:random}

Next, we formulate a stochastic scheme for the (random) step position variable, $X(t)$, coupled with the (random) number, $\varrho_j(t)$, of adatoms at lattice site $j$ ($j=0,\,1,\ldots\,, N-1$).
We discretize time, $t=t_n$, with a constant, sufficiently small timestep, $\tau=t_{n+1}-t_n$. 
The main idea is to describe how $X(t)$ changes at each time increment by relating $X(t_{n+1})$ to $X(t_n)$ via the values of a random variable, $\xi(t_n)$; see Fig.~\ref{fig:kinetics}. 
For ease in notation, set $q^n=q(t_n)$, $\varrho_j^n=\varrho_j(t_n)$, $\xi^n=\xi(t_n)$ and $X^n=X(t_n)$.

Consider the discrete random variable $\xi(t)$ that takes values in the set $\{-1, -2, 0, 1, 2\}$. These integer values correspond to the possible atomistic events at $t=t_n$ (Fig.~\ref{fig:kinetics}).
In particular, $\xi^n=1$ or $2$ if an adatom attaches to the step from the upper or lower terrace, respectively, so that the step advances; and $\xi^n=-1$ or $-2$ if the edge atom detaches toward the upper or lower terrace, so that the step retreats. The value $\xi^n=0$
amounts to processes that do not cause step motion for $t_n\le t<t_{n+1}$.

The microscale step position is updated with time as
\begin{equation}\label{eq:step-micro}
X^{n+1}=\left\{\begin{array}{ll} X^n & \mbox{if}\ \xi^n=0,\cr
                                                          X^n+a & \mbox{if}\ \xi^n=1\ \mbox{or}\ 2,\cr
                                                          X^n-a& \mbox{if}\ \xi^n=-1\ \mbox{or}\ -2.
                           \end{array}\right.
\end{equation}
We supplement this rule with the probabilities
\begin{align}\label{eq:probab}
&\P(\xi^n=1)=D\phi_-\tau \varrho^n_{q^n-1}, \ \P(\xi^n=2)=D\phi_+\tau\varrho^n_{q^n+1},\nonumber\\
&\P(\xi^n=-1)=Dk\phi_-\tau,\ \P(\xi^n=-2)=Dk\phi_+\tau,
\end{align}
which express rules (iv) and (v) of Sec.~\ref{subsec:kins} [cf. Fig.~\ref{fig:kinetics}(c)--(f)]. For example, $\P(\xi^n=1)$ is the probability that an adatom attaches to the step edge
from the upper terrace, depending on the adatom number, $\varrho_j^n$, at the site left of the edge, $j=q^n-1$. Clearly, $\P(\xi^n=0)$ follows from Eq.~\eqref{eq:probab}: $\P(\xi^n=0)=1-D\tau(\phi_-\varrho^n_{q^n-1}+\phi_+\varrho^n_{q^n+1})-D\tau k(\phi_-+\phi_+)$. 

It remains to prescribe the discrete scheme for the adatom number, $\varrho_j^n$, per lattice site. For sites sufficiently away from the step edge, we have
\begin{align}\label{eq:rho_away-step}
\varrho_j^{n+1}&=(1-2D\tau)\varrho_j^n+D\tau(\varrho_{j-1}^n+\varrho_{j+1}^n)-\frac{\tau}{\ta}\varrho_j^n+\tau f,\nonumber\\
&\qquad j\neq 0,\,q^n-2,\,q^n-1,\,q^n,\,q^n+1,\,N-1,
\end{align}
which expresses the usual unbiased random walk on a lattice [Fig.~\ref{fig:kinetics}(a), (b)] in the presence of desorption and external material deposition from above.
At the domain boundaries ($j=0,\, N-1$), for definiteness we impose
\begin{subequations}\label{eq:rho-bdry}
\begin{align}
\varrho_0^{n+1}&=(1-D\tau)\varrho_0^n+D\tau \varrho_1^n+f_{\rm in}\tau-\frac{\tau}{\ta}\varrho_0^n+\tau f,\label{eq:rho-bdry-0}\\
\varrho_{N-1}^{n+1}&=(1-2D\tau)\varrho_{N-1}^n+D\tau\varrho_{N-2}^n-\frac{\tau}{\ta}\varrho_{N-1}^{n}+\tau f.\label{eq:rho-bdry-1}
\end{align}
\end{subequations}
 Alternatively, one can impose relations that amount to screw-periodic boundary conditions for $\varrho_j$. 
For the remaining sites, the scheme accounts for atom attachment/detachment at the step edge:
\begin{subequations}\label{eq:rho-edge}
\begin{align}
\varrho_{q^n-2}^{n+1}&=(1-2D\tau)\varrho_{q^n-2}^n+D\tau (\varrho^n_{q^n-3}+\varrho_{q^n-1}^n)\nonumber\\
&\qquad -\frac{\tau}{\ta}\varrho_{q^n-2}+f\tau+\1(\xi^n=-1),\label{eq:rho-edge1}\\
\varrho_{q^n-1}^{n+1}&=(1-D\tau)\varrho_{q^n-1}^n+D\tau \varrho_{q^n-2}^n-\frac{\tau}{\ta}\varrho_{q^n-1}^n+f\tau\nonumber\\
&\qquad -\1(\xi^n=1),\label{eq:rho-edge2}\\
\varrho_{q^n}^{n+1}&=(1-D\tau)\varrho_{q^n}^n+D\tau\varrho_{q^n+1}^n-\frac{\tau}{\ta}\varrho_{q^n}^n+f\tau\nonumber\\
&\qquad +\1(\xi^n=-2),\label{eq:rho-edge3}\\
\varrho_{q^n+1}^{n+1}&=(1-2D\tau)\varrho_{q^n+1}^n+D\tau(\varrho_{q^n}^n+\varrho_{q^n+2}^n)\nonumber\\
&\qquad -\frac{\tau}{\ta}\varrho_{q^n+1}^n+f\tau-\1(\xi^n=2).\label{eq:rho-edge4}
\end{align}
\end{subequations}
In the above, $\1(\cdot)$ is the indicator function, viz., $\1(\mbox{A})=1$ if the event A occurs, and $\1(\mbox{A})=0$ otherwise. Thus, the presence of this indicator in Eqs.~\eqref{eq:rho-edge} signifies the addition or removal of an adatom to/from the corresponding lattice site when the step edge moves  [Fig.~\ref{fig:kinetics}(c)--(f)].

Some remarks on the meaning of Eqs.~\eqref{eq:rho-edge} are in order; see also Fig.~\ref{fig:kinetics}. By Eqs.~\eqref{eq:rho-edge1} and~\eqref{eq:rho-edge4} pertaining to sites $q^n-2$ and $q^n+1$, an adatom at these sites can either hop to or from any of the two adjacent sites with rate $D$ [Figs.~\ref{fig:kinetics}(a), (b)]; or evaporate with rate $\ta^{-1}$; or be deposited from the vapor to the surface with rate $f$; or come from an atom detaching from the step edge [Eq.~\eqref{eq:rho-edge1} and Fig.~\ref{fig:kinetics}(e)]; or attach to the step [Eq.~\eqref{eq:rho-edge4} and Fig.~\ref{fig:kinetics}(d)]. In the same vein, in regard to Eqs.~\eqref{eq:rho-edge2} and~\eqref{eq:rho-edge3} for sites $q^n-1$ and $q^n$, an adatom at these sites can either hop to or from the adjacent site of the same terrace with rate $D$;  or evaporate; or be deposited from above; or attach to the step edge [Eq.~\eqref{eq:rho-edge2} and Fig.~\ref{fig:kinetics}(c)]; or come from the detachment of the edge atom [Eq.~\eqref{eq:rho-edge3} and Fig.~\ref{fig:kinetics}(f)].

Note that an atomistic process amounting to step permeability at the BCF level~\citep{OzdemirZangwill92} can plausibly be incorporated, in an ad hoc fashion, into our scheme. A particular choice for such a process is that an adatom directly hopes from a site, say, $q^n-1$, in the upper terrace to $q^n$ in the lower terrace, and vice versa, without attaching to the step. The respective probability is considered as proportional to: an extra, appropriately scaled, permeability rate; and the difference of adatom numbers at the two relevant sites. We do not further pursue this extension in our analysis.

\subsection{Averaging of stochastic scheme}
\label{subsec:mean}

We now average out the governing stochastic laws of Sec.~\ref{subsec:random} in the limit $\tau\to 0$ by keeping the spacing $a$ fixed.
To simplify the analysis, we henceforth apply the condition that $a\ll
1$ and take into account that, as argued in Sec.~\ref{subsec:noise}  for the appropriate scaling of parameters,  the
stochastic noise for the step position, $X(t)$, is negligible for
small enough $a$. Therefore, we carry out the averaging procedure by allowing the mean of $\varrho_q(t)$, where $q$ is a stochastic variable, to
be set approximately equal to $\bar{\varrho}_{\bar q}(t)$~\cite{comment1}. 

By Eqs.~\eqref{eq:step-micro} and~\eqref{eq:probab}, we obtain the expectation
\begin{align}\label{eq:X-ave}
&\E[X^{n+1}-X^n]=a\{\P(\xi^n=1\ \mbox{or}\ 2)-\P(\xi^n=-1\ \mbox{or}\ -2)\}\nonumber\\
 &\qquad \approx Da\{(\phi_-\bar\varrho_{\bar q^n-1}^n+\phi_+\bar\varrho_{\bar q^n+1}^n)-k(\phi_-+\phi_+)\}\tau,
\end{align}
where $\E[X]\equiv \bar X$.
We also compute the variance 
\begin{align}\label{eq:X-var}
\V[X^{n+1}-X^n]&= Da^2\{(\phi_-\bar\rho_{\bar q-1}+\phi_+\bar\rho_{\bar q+1})\nonumber\\
&\qquad +k(\phi_-+\phi_+)\}\tau+\mathcal O(\tau^2).
\end{align}

In the limit $\tau\to 0$, we thus derive a mean step velocity law at $t=t_n$ in terms of $\bar\varrho_j^n$ where $j$ denotes sites adjacent to the step edge: 
\begin{align}
\frac{dx_{\rm s}}{dt}&\equiv \lim_{\tau\to 0}\E\Biggl[\frac{X^{n+1}-X^n}{\tau}\Biggr]\nonumber\\
&\approx a\{D(\phi_-\bar\varrho_{\bar q-1}+\phi_+\bar\varrho_{\bar q+1})-Dk(\phi_-+\phi_+)\};\label{eq:step-vel-micro}
\end{align}
here, $\bar q=\bar q(t)$ and $x_{\rm s}(t)=\bar X(t)=\bar q(t)a$ denote averages.

Accordingly, as $\tau\to 0$ the heuristic limit of the mean of Eqs.~\eqref{eq:rho_away-step} and~\eqref{eq:rho-bdry}, for $j\neq q-2, q-1, q, q+1$,  reads
\begin{equation}\label{eq:rho_away-step-taulimit}
\frac{d\bar\varrho_j}{dt}=D(\bar\varrho_{j-1}-2\bar\varrho_j+\bar\varrho_{j+1})-\frac{1}{\ta}\bar\varrho_j+f;
\end{equation}
\begin{subequations}\label{eq:rho-bdry-taulimit}
\begin{align}
\frac{d\bar\varrho_0}{dt}&=D(\bar\varrho_1-\bar\varrho_0)+f_{\rm in}-\frac{1}{\ta}\bar\varrho_0+f,\\
\frac{d\bar\varrho_{N-1}}{dt}&=-2D\bar\varrho_{N-1}+D\bar\varrho_{N-2}-\frac{1}{\ta}\bar\varrho_{N-1}+f.
\end{align}
\end{subequations}
For lattice sites near the step edge, the scheme reads
\begin{subequations}\label{eq:rho-edge-taulimit}
\begin{align}
\frac{\partial\bar\varrho_{\bar q-2}}{\partial t}&=D(\bar\varrho_{\bar q-3}-2\bar\varrho_{\bar q-2}+\bar\varrho_{\bar q-1})-\frac{1}{\ta}\bar\varrho_{\bar q-2}\nonumber\\
&\qquad +f+Dk\phi_-,\\
\frac{\partial\bar\varrho_{\bar q-1}}{\partial t}&=D(\bar\varrho_{\bar q-2}-\bar\varrho_{\bar q-1})-\frac{1}{\ta}\bar\varrho_{\bar q-1}+f\nonumber\\
&\qquad -D\phi_-\bar\varrho_{\bar q-1},\\
\frac{\partial\bar\varrho_{\bar q}}{\partial t}&=D(\bar\varrho_{\bar q+1}-\bar\varrho_{\bar q})-\frac{1}{\ta}\bar\varrho_{\bar q}+f+Dk\phi_+,\\
\frac{\partial\bar\varrho_{\bar q+1}}{\partial t}&=D(\bar\varrho_{\bar q+2}-2\bar\varrho_{\bar q+1}+\bar\varrho_{\bar q})-\frac{1}{\ta}\bar\varrho_{\bar q+1}+f\nonumber\\
&\qquad -D\phi_+\bar\varrho_{\bar q+1}.
\end{align}
\end{subequations}
Equations~\eqref{eq:rho_away-step-taulimit} and~\eqref{eq:rho-edge-taulimit} are recast into the compact form
\begin{align}\label{eq:rho_away-step-comp}
\frac{\partial\bar\varrho_j}{\partial t}&=D(\bar\varrho_{j-1}-2\bar\varrho_j+\bar\varrho_{j+1})-\frac{1}{\ta}\bar\varrho_j+f \nonumber\\
&  +Dk\phi_- \delta_{j,\bar q-2}+[D(\bar\varrho_{\bar q-1}-\bar\varrho_{\bar q})-D\phi_-\bar\varrho_{\bar q-1}]\delta_{j,\bar q-1}\nonumber\\
&  +[D(\bar\varrho_{\bar q}-\bar\varrho_{\bar q-1})+Dk\phi_+]\delta_{j,\bar q}-D\phi_+\bar\varrho_{\bar q+1}\delta_{j,\bar q+1},
\end{align}
in which $j\neq 0,\, N-1$ and $\delta_{i,j}$ denotes Kronecker's delta.

Equation~\eqref{eq:step-vel-micro} couples the discrete mean step velocity law with the average adatom numbers on each side of the step. In the limit $a\to 0$, this coupling will give rise to a mass conservation statement involving the values of the adatom flux directly to the left and right of the edge (Sec.~\ref{sec:limit}). This flux can be determined via Eqs.~\eqref{eq:rho_away-step-taulimit}--\eqref{eq:rho-edge-taulimit}. 
To reduce the discrete equations to BCF-type laws, we need to appropriately scale variables and parameters 
with the system size, $N=a^{-1}$ (Sec.~\ref{sec:limit}).

\section{Scaling limit as $a\to 0$}
\label{sec:limit}

Next, we carry out the scaling limit of Eqs.~\eqref{eq:step-vel-micro}--\eqref{eq:rho-edge-taulimit} as $a\to 0$ by use of Eq.~\eqref{eq:rho_away-step-comp}. For this purpose, we restrict attention
to macroscopic times by defining
\begin{equation}
\tilde t=at,\quad
{\tilde \tau}_{\rm e}=a\ta,\label{eq:macro-time}
\end{equation}
and the variable
\begin{equation}
\tilde\rho_j(\tilde t)=\bar\varrho_j(t)/a,\label{eq:ad-density-discr}
\end{equation}
which is the adatom number density.
We also consider $\tilde t, \tilde \ta=\mathcal O(1)$  and $\tilde\rho_j=\mathcal O(1)$ (bounded). 
For notational economy, we will drop the tildes and also replace $\bar q$ by $q$.

By Eq.~\eqref{eq:step-vel-micro}, the mean step velocity law reads
\begin{equation}\label{eq:step-vel-macro}
\frac{dx_{\rm s}}{dt}=(r_{\rm a}^- \rho_{q-1}-r_{\rm d}^-)+(r_{\rm a}^+ \rho_{q+1}-r_{\rm d}^+),
\end{equation}
where both sides are bounded as $a\to 0$. 
Equation~\eqref{eq:step-vel-macro} forms the core of our scaling argument. The requisite kinetic coefficients are defined by
\begin{equation}\label{eq:lim-parmts}
r^{\pm}_{\rm d}=Dk\phi_{\pm},\quad r^{\pm}_{\rm a}=D\phi_{\pm}a,
\end{equation}
which are the mesoscopic detachment (d) and attachment (a) rates to the left ($-$) or right ($+$) of the step edge. Hence,  in order to obtain a linear kinetic relation for the adatom flux at the step edge,  it is reasonable to assume that, as $a\to 0$, the rates of Eqs.~\eqref{eq:lim-parmts} are finite and independent of $a$ (Sec.~\ref{subsec:rates-sc}).

Thus, Eq.~\eqref{eq:rho_away-step-comp} for the adatom number density becomes
\begin{align}\label{eq:rho_comp-sc}
a\frac{\partial \rho_j}{\partial t}&=D(\rho_{j-1}-2\rho_j+\rho_{j+1})-\frac{1}{\ta}\rho_j+f a^{-1} \nonumber\\
& +\{Da(\rho_q-\rho_{q-1})(\delta_{j,q}-\delta_{j,q-1})\nonumber\\
& +(r_{\rm d}^+\delta_{j,q}-r_{\rm a}^+\rho_{q+1}\delta_{j,q+1})\nonumber\\
& +(r_{\rm d}^-\delta_{j,q-2}-r_{\rm a}^-\rho_{q-1}\delta_{j,q+1})\}a^{-1},
\end{align}
where $j=1,\,\ldots\,,N-2$. In the following, we use Eq.~\eqref{eq:rho_comp-sc} in order to express the step velocity, the right-hand side of Eq.~\eqref{eq:step-vel-macro}, as the sum of adatom fluxes toward the step edge. We additionally impose Eqs.~\eqref{eq:rho-bdry-taulimit}, suitably scaled, at the domain boundaries, $x=0,\,1$.  Alternatively, we can impose screw-periodic boundary conditions. 

\subsection{Scaling of atomistic rates}
\label{subsec:rates-sc}

We now discuss the scaling of the kinetic parameters with $a$, by inspection of Eqs.~\eqref{eq:step-vel-macro}--\eqref{eq:rho_comp-sc}. First, we set
\begin{equation}
\mathfrak D\equiv Da^2=\mathcal O(1), \label{eq:D-sc}
\end{equation}
i.e., require that the rate $D$ scale with the system size as $1/a^2=N^2$. This $\mathfrak D$ expresses the usual macroscopic diffusivity resulting from a random walk on a lattice~\cite{Chorin-book}. By Eqs.~\eqref{eq:rho-bdry-taulimit}, \eqref{eq:lim-parmts} and~\eqref{eq:rho_comp-sc}, we also assume that
\begin{equation}
\phi_{\pm}=\mathcal O(a),\ k=\mathcal O(a),\ f=\mathcal O(a),\ f_{\rm in}=\mathcal O(1),\label{eq:phi-k-sc}
\end{equation}
and define
\begin{equation}\label{eq:f-def}
F\equiv f a^{-1}=\mathcal O(1).
\end{equation}

Equations~\eqref{eq:D-sc} and~\eqref{eq:phi-k-sc}  are in agreement with assumptions made in previously published results, e.g., \cite{PatroneEM14,PatroneM14},  for the corresponding kinetic regime,  and suffice for deriving a kinetic relation for the adatom flux
as a linear function of the adatom density at the step edge. The parameters $\mD^{-1} f_{\rm in}$ and $\mD^{-1} F$ should be sufficiently small, consistent with the diluteness hypothesis.

 We alert the reader that Eqs.~\eqref{eq:phi-k-sc} preclude
$\phi_+\approx 1$ or $\phi_-\approx 1$. This case signifies
the absence of an Ehrlich-Schwoebel
barrier~\cite{PimpinelliVillain99,ES}.  In this regime, the dominant
balance of terms in the averaged microscopic description yields, to leading order in $a$, an anticipated Dirichlet boundary condition: the continuum-scale adatom density, $\rho$, equals an equilibrium density at the step edge~\cite{PimpinelliVillain99,BCF51,PatroneEM14}. 
Indeed, in this case the expression in the third or fourth line of Eq.~\eqref{eq:rho_comp-sc} is large, $\mathcal O(1/a)$, and, thus, should vanish to leading order. This amounts to a step-edge adatom density $\rho^{\pm}=\lim_{a\to 0}\rho_{q\pm 1}=r_{\rm d}^+/r_{\rm a}^+=r_{\rm d}^-/r_{\rm a}^-$ to the right ($+$) or left ($-$)
of the step edge; cf.~Eq.~\eqref{eq:rho-eq-def}. Moreover, the right-hand side of mean step
velocity law~\eqref{eq:step-vel-macro} converges to a BCF-type law, with the velocity determined by the derivative of the adatom density
on the corresponding side. Hence, formally, we still arrive at 
BCF-type laws albeit with a Dirichlet boundary condition, to be contrasted with the linear relation for the adatom flux in Eqs.~\eqref{eq:bdry-Robin} below. However, unlike the
case $\phi_+, \phi_- = \mathcal O(a)$, the variance of
the step position does not vanish as $a \to 0$ when $\phi_+$ or $\phi_-$ is $\mathcal O(1)$; cf. Eq.~\eqref{eq:X-SDE} below. Hence, the law of large numbers becomes questionable. 
We leave a more systematic treatment of this regime to near-future work. In the following analysis, we assume that Eqs.~\eqref{eq:phi-k-sc} hold,
unless stated otherwise.

\subsection{On limit of stochastic noise}
\label{subsec:noise}

Next, we show that the stochastic noise underlying mean step velocity law~\eqref{eq:step-vel-macro} vanishes as $a\to 0$. By Eqs.~\eqref{eq:X-ave} and~\eqref{eq:X-var}, the stochastic differential equation for the step position variable, $X(t)$,  is 
\begin{equation}
dX_t\approx c_{\rm s} \,dt+\sqrt{a}c_{\rm n}\,dW_t,\label{eq:X-SDE}
\end{equation}
where $dt=t_{n+1}-t_n=\tau$, $dX_t=X^{n+1}-X^n$ and $W_t$ is the Wiener process~\cite{Chorin-book} so that $dW_t=W^{n+1}-W^n$ is discrete ``white noise''. The finite quantities $c_{\rm s}$ and $c_{\rm n}$ come from the expectation
$\E[dX_t]$ [Eq.~\eqref{eq:X-ave}] and standard deviation $\sqrt{\V[dX_t]}$ [Eq.~\eqref{eq:X-var}] of $X^{n+1}-X^n$ as $dt \to 0$: 
\begin{equation*}
c_{\rm s}=r_{\rm a}^- \rho_{q-1}-r_{\rm d}^-+r_{\rm a}^+ \rho_{q+1}-r_{\rm d}^+,
\end{equation*}
\begin{equation}
c_{\rm n}=(r_{\rm a}^-\rho_{q-1}+r_{\rm d}^-+r_{\rm a}^+\rho_{q+1}+r_{\rm d}^+)^{1/2}.
\end{equation}

By inspection of Eq.~\eqref{eq:X-SDE} under Eqs.~\eqref{eq:D-sc} and~\eqref{eq:phi-k-sc}, the white noise vanishes as $a\to 0$ provided the densities $\rho_{q-1}$ and $\rho_{q+1}$ approach finite values. This is not surprising: the step front can only move by distance $\pm a$ each time which in turn causes a negligibly small variance of its random motion. 
Hence, in this regime, step motion can be viewed as a phenomenon in the context of the law of large numbers. It should be noted, however, that a mesoscale description in which the noise is preserved as $a\to 0$ may result under different kinetics or scaling scenario.  It is worthwhile observing, for example, that in the absence of an Ehrlich-Schwoebel barrier~\cite{ES}, when $\phi_+\approx 1$ or $\phi_-\approx 1$, the coefficient of $dW_t$ in Eq.~\eqref{eq:X-SDE} becomes $\mathcal O(1)$.  This issue deserves to be the subject of future studies.

\subsection{Step flow limit}
\label{subsec:step-lim}

We now complement Eq.~\eqref{eq:step-vel-macro} with a description of the adatom number density, $\rho_j(t)$, as $a\to 0$. Suppose the step position is still denoted $x_{\rm s}(t)$ in this limit. By slightly abusing notation, we replace $\rho_j(t)$ by the function $\rho(t,x)$, assuming that this limit exists; $0<x<1$ with $x\neq x_{\rm s}(t)$ and $t>0$. Furthermore, $\mathfrak D$, $r_{\rm d}^{\pm}$, $r_{\rm a}^\pm$, $\ta$, and $F$ take their finite limiting values. We will suppress the time dependence of $\rho(t,x)$ for algebraic convenience.

Consider Eq.~\eqref{eq:rho_comp-sc}. First, $a(\partial \rho_j/\partial t)\approx a[\partial\rho(t,x)/\partial t]\to 0$ for fixed time $t$, since $\partial\rho_j/\partial t$ is bounded. Second, it is tempting to replace the second-order difference term, $a^{-2}(\rho_{j+1}-2\rho_j+\rho_{j-1})$, by the Laplacian of $\rho(x)$, $\Delta_x\rho$, for $x<x_{\rm st}$ and $x>x_{\rm st}$. A word of caution is in order. If $j=q-1$ or $j=q$, the above discrete term involves values of $\rho_j$ on both sides of the step edge; however, $\rho(x)$ can be discontinuous across the step. In an effort to describe the limit of Eq.~\eqref{eq:rho_comp-sc} transparently, we introduce reference densities $\rho_{\rm s}^\pm$ such that the scheme for the adatom number density at sites adjacent to the step edge reads ~\cite{PimpinelliVillain99,PatroneM14}
\begin{subequations}\label{eq:discrete-mod}
\begin{align}
j=q-1:\ 0&=\mD [a^{-2}(\rho_{j-1}-2\rho_{j}+\rho_{\rm s}^-)]-\frac{1}{\ta}\rho_{j}+F\nonumber\\
& -\{r_{\rm a}^-\rho_{j}+\mD [a^{-1}(\rho_{\rm s}^--\rho_{j})]\}a^{-1},\label{eq:discrete-mod1}\\
j=q:\ 0&=\mD[a^{-2}[(\rho_{j+1}-2\rho_j+\rho_{\rm s}^+)]-\frac{1}{\ta}\rho_j+F\nonumber\\
& +\{r_{\rm d}^+  +\mD[a^{-1}(\rho_j-\rho_{\rm s}^+)]\}a^{-1}.\label{eq:discrete-mod2}
\end{align}
\end{subequations}
The densities $\rho_{\rm s}^{\pm}$ can be thought of as representing the continuum limits of $\rho_j$ at either side of the step edge, and can be determined so that
they produce the appropriate adatom fluxes to the right ($+$) or left ($-$) of the step. Specifically, $\pm a^{-1}(\rho_j-\rho_{\rm st}^\pm)$ is let to approach
$(\partial\rho/\partial x)^\pm$, the respective value of the derivative of $\rho(x)$, for $j=q$ ($+$) or $j=q-1$ ($-$). These terms contribute to the desired boundary conditions as shown below.

In the limit $a\to 0$, Eq.~\eqref{eq:rho_comp-sc} becomes
\begin{align}\label{eq:adatom-law}
0&=\{{\mathfrak D}\Delta_x\rho-\ta^{-1}\rho(x)+F\}[\theta(x-x_{\rm s})+\theta(x_{\rm s}-x)]\nonumber\\
&\qquad +\delta_{x_{\rm s}}^+(-\mathcal J^++r_{\rm d}^+-r_{\rm a}^+\rho^+)\nonumber\\
&\qquad +\delta_{x_{\rm s}}^- (\mathcal J^-+r_{\rm d}^--r_{\rm a}^-\rho^-),\qquad 0<x<1.
\end{align}
In the above, $\theta(x)$ is the Heaviside function [$\theta(x)=0$ if $x<0$ and $\theta(x)=1$ if $x>0$]; $\delta_{x_{\rm s}}^\pm=\lim_{a\to 0}(a^{-1}\delta_{j,l})$ is the delta function centered at $x_{\rm s}$ to the left ($-$) or right ($+$) of the step edge, for $l=q-2,\,q-1$ and $l=q, q+1$, respectively; and $\mathcal J^\pm$ is the adatom flux restricted at the step edge, viz., 
\begin{align}
\mathcal J^+&=-\mathfrak D \left(\frac{\partial\rho}{\partial x}\right)^+=-\mathfrak D \lim_{a\to 0}\left(\frac{\rho_q-\rho_{\rm s}^+}{a}\right),\nonumber\\
\mathcal J^-&=-\mathfrak D\left(\frac{\partial\rho}{\partial x}\right)^-=-\mathfrak D \lim_{a\to 0}\left(\frac{\rho_{\rm s}^--\rho_{q-1}}{a}\right).\label{eq:flux-}
\end{align}
Evidently, there is no convective term present in $\mathcal J^\pm$, which is consistent with the elimination of $\partial\rho/\partial t$. This feature signifies the quasistatic regime.

Equation~\eqref{eq:adatom-law} is equivalent to a diffusion equation on each terrace along with kinetic boundary conditions involving the adatom flux at the step edge:
\begin{equation}\label{eq:diff}
\mD\Delta_x\rho-\ta^{-1}\rho(x)+F=0,\quad  x\neq x_{\rm s},
\end{equation}
\begin{align}\label{eq:bdry-Robin}
\mathcal J^+&=-r_{\rm a}^+(\rho^+-r_{\rm d}^+/r_{\rm a}^+),\qquad x=x_{\rm s}^+,\nonumber\\
\mathcal J^-&=r_{\rm a}^-(\rho^--r_{\rm d}^-/r_{\rm a}^-),\qquad x=x_{\rm s}^-,
\end{align}
where
\begin{equation}\label{eq:rho-eq-def}
\frac{r_{\rm d}^+}{r_{\rm a}^+}=\frac{r_{\rm d}^-}{r_{\rm a}^-}\equiv \rho_{\rm eq}=\lim_{a\to 0}\left(\frac{k}{a}\right),
\end{equation}
which is finite by Eq.~\eqref{eq:phi-k-sc}. This $\rho_{\rm eq}$ represents the equilibrium number density of adatoms at the step edge; cf.~\cite{PimpinelliVillain99,PatroneM14,Zhao05}.
Thus, step velocity law~\eqref{eq:step-vel-macro} reads
\begin{align}\label{eq:step-vel-scaled}
\frac{dx_{\rm s}}{dt}&=r_{\rm a}^-(\rho^--\rho_{\rm eq})+r_{\rm a}^+(\rho^+-\rho_{\rm eq})=\mathcal J^--\mathcal J^+.
\end{align}
Equations~\eqref{eq:diff}, \eqref{eq:bdry-Robin} and~\eqref{eq:step-vel-scaled} are the desired BCF-type laws.  Notably, in the regime where $r_{\rm a}^+$ or $r_{\rm a}^-$ becomes large, but $\rho_{\rm eq}$ and $\mathcal J^{\pm}$ remain bounded, Eqs.~\eqref{eq:bdry-Robin} formally give rise to a Dirichlet boundary condition~\cite{PimpinelliVillain99}.  

Finally, we need to add conditions at the domain boundaries, $x=0$ and $1$. By Eqs.~\eqref{eq:rho-bdry-taulimit}, we obtain 
\begin{align}\label{eq:dom-cond0}
0&=\lim_{a\to 0}\{\mD [a^{-1}(\rho_1-\rho_0)]+f_{\rm in}-a \ta^{-1}\rho_0+aF\}\nonumber\\
  &\Rightarrow \mathcal J(0)=-\mD \left(\frac{\partial\rho}{\partial x}\right)\Biggl|_{x=0}=f_{\rm in},
\end{align}
\begin{align}\label{eq:dom-cond1}
0&=\lim_{a\to 0}\{-\mD\rho_{N-1}-\mD(\rho_{N-1}-\rho_{N-2})\}\nonumber\\
 &\Rightarrow \rho(1)=0.
\end{align}
 Alternatively, screw-periodic boundary conditions on $\rho$ can be imposed.

\section{Discussion}
\label{sec:discussion}
In this section, we briefly discuss issues that underlie the exposition and formal analysis of Secs.~\ref{sec:model} and~\ref{sec:limit}. 

\subsection{Convergence of atomistic scheme}
\label{subsec:convergence}

 Thus far, we have provided a derivation of the BCF-type model from an
atomistic scheme based on heuristic asymptotics. To make the
derivation mathematically rigorous, it is useful to make the analogy of
the atomistic dynamics to a finite-difference numerical scheme approximating the
continuous description of the BCF-type model. The lattice parameter $a$, which approaches zero, is
identified with the mesh size of the discretization.

Let us briefly sketch the main ideas of the proof of convergence of the numerical scheme to the BCF-type model; the details lie beyond the scope of this paper. As usual,
the convergence of the scheme involves both consistency and stability
analysis. The consistency for the scheme essentially follows the
heuristic asymptotic arguments provided above in the derivation. The
stability is more subtle. A difficulty comes from the quasistatic
time scaling on the left-hand side of Eq.~\eqref{eq:rho_comp-sc}: The
small parameter $a$ multiplying the time derivative of $\rho$ requires stability
for effectively long time evolution. Hence, an energy estimate is
needed to show that the discrete system is dissipative. This
amounts to establishing a gradient flow structure for the atomistic
scheme, which is expected to be similar to that on the continuous scale for the BCF-type system with detailed balance \cite{Otto}.

\subsection{Multiple steps}
\label{subsec:m-steps}

Our analysis can be extended to more than one non-interacting, ordered steps without difficulty. The main observation is that the above derivation of step motion laws is local, based on local atomistic laws. Specifically, 
boundary conditions \eqref{eq:bdry-Robin} and step velocity law \eqref{eq:step-vel-scaled} both result from the mass exchange between the edge atom and adatoms in the neighboring lattice sites.
Hence, the derivation of mesoscale laws for
a monotone step train follows directly, provided the steps do not interact elastically and are sufficiently far apart. 
For example, if the system consists of $M$ non-interacting steps with the same kinetic rates everywhere, the number of adatoms at $t=t_{n+1}$ at the $k$th step edge, which is at site $q_k(t)$ ($k=1,\,2,\,\ldots M$), is given by 
\begin{align}\label{eq:rho-edge-k}
\varrho_{q^n_k}^{n+1}&=(1-D\tau)\varrho_{q^n_k}^n+D\tau\varrho_{q^n_k+1}^n-\frac{\tau}{\ta}\varrho_{q^n_k}^n+f\tau\nonumber\\
&\qquad +\1(\xi^n_k=-2),
\end{align} 
where the random variable $\xi_k(t)$ indicates the atomistic events relevant to the $k$th step; cf. Eqs.~\eqref{eq:rho-edge}. The local probabilistic rules for $\xi_k$ are dictated by Eqs.~\eqref{eq:probab}.

In this case, in the scaling limit each step moves according to velocity law
\eqref{eq:step-vel-scaled} with the adatom density determined by
quasistatic diffusion on each terrace with the same boundary conditions at each step edge.
However, as our atomistic model does not include elastic response of
the lattice, the system of multiple steps is deemed as physically incomplete. It is an interesting and challenging research
direction to understand the elastic interaction between multiple steps
starting from atomistic models.

\section{Conclusion}
\label{sec:conclusion}

We formally derived a set of quasistatic motion laws for non-interacting steps in 1D, starting with a stochastic scheme for the hopping of atoms on a  lattice. The derived laws form the core of known BCF-type theories.  Our scheme was adopted on the basis of a kinetic restricted SOS model for a dilute system of adatoms. By our methodology, the step edge is treated as a front that propagates via the attachment/detachment of atoms. This process is described by a random variable that takes values under probabilistic rules associated with step kinetics.  To the best of our knowledge, our approach, based on the systematic averaging of a stochastic scheme, has not been previously applied in epitaxial growth. 

Our formal analysis reveals some key features of the passage from atomistic rules to mesoscale laws for line defects in 1+1 dimensions. The emergence of BCF-type laws, including the full boundary conditions for the adatom density at the step edge, is intimately connected to certain,  previously known,  scalings of the time variable and the atomistic rates with the system size, $N=a^{-1}$ (Sec.~\ref{sec:limit}). Our present approach  firmly places these scalings in the context of a stochastic scheme,  unveiling a particular dominant balance for the adatom density and flux as the lattice spacing, $a$, approaches zero. Our analysis also  describes the variance of the stochastic step fluctuations in this limit.  In particular, we show that the stochastic noise vanishes in this limit  when step-edge barriers are present on both sides of the step.

Our work points to several pending issues. An issue is the possible
emergence from atomistic rules of a {\em stochastic} mesoscale model,
in which the noise plays a significant role as $a\to 0$. Furthermore,
in experimental situations, steps interact as force dipoles in
homoepitaxy and force monopoles or otherwise in hereroepitaxy. Hence,
our current treatment needs to include elastic effects by taking into
account the strain dependence of kinetic rates. Lastly, the derivation
of a BCF-type description in 2D, where steps meander in the presence
of kinks~\cite{Shitara93}, and islands form, is a viable direction of
future research.

\section*{ACKNOWLEDGMENTS}
We wish to thank Professor R.~E. Caflisch, Professor T.~L. Einstein,
Professor R.~V. Kohn, Dr. P.~N. Patrone, and Professor A. Pimpinelli
for valuable discussions.  JL was supported in part by the Alfred
P. Sloan Fellowship and the NSF via Grant No. DMS-1312659. DM was
supported by NSF via Grant No. DMS 08-47587. JGL and DM were also
supported by the NSF Research Network Grant No. RNMS11-07444 (KI-Net)
in the fall of 2013 when part of this work was initiated.


\begin{thebibliography}{[30]}
\bibitem{comment1}This procedure reasonably assumes that the operation $\E[\cdot]$ commutes with the limit $a\to 0$.
\end{thebibliography}
\end{document}